\documentclass[aps,prl,superscriptaddress,showpacs,twocolumn]{revtex4}
\usepackage{amsfonts}
\usepackage{amsmath}
\usepackage{amssymb}
\usepackage{graphicx}

\begin{document}

\title{Vertical absorption edge and universal onset conductance in
semi-hydrogenated graphene}
\author{Lei Chen$^{(a)}$}
\affiliation{School of Physics, Peking University, Beijing 100871, China}
\author{Zhongshui Ma}
\affiliation{School of Physics, Peking University, Beijing 100871, China}
\author{C. Zhang$^{(b)}$}
\affiliation{School of Engineering Physics, University of Wollongong, New South Wales
2552, Australia}

\begin{abstract}
We show that for graphene with any finite difference in the on-site energy
between the two sub-lattices ($\Delta$), The optical absorption edge is
determined by the $\Delta$. The universal conductance will be broken and the
conductance near the band edge varies with frequency as $1/\omega^2$.
Moreover, we have identified another universal conductance for such systems
without inversion symmetry, i.e., the onset conductance at the band edge is $%
\sigma_c=2\sigma_0=\pi e^2/2h$, independent of the size of the band gap. The
total integrated optical response is nearly conserved despite of the opening
of the band gap.
\end{abstract}

\maketitle

In recent years, graphene has attracted a great deal of interest\cite%
{novo1,novo2,zhang1,berg}. New physics have been predicted and observed,
such as electron-hole symmetry and half-integer quantum Hall effect\cite%
{novo2,zhang1}, finite conductivity at zero charge-carrier concentration\cite%
{novo2}, and the strong suppression of weak localization\cite%
{suzuura,morozov,khveshchenko}. In graphene, the conduction and valence
bands touch each other at six equivalent points, the $K$ and $K^{\prime}$
points in the Brillouin zone. Near these points the electrons behave like
massless Dirac Fermions. One of the most striking features of the massless
Dirac Fermion is that the optical conductance is a universal constant, $%
\sigma_0=\pi e^2/4h$. This was calculated theoretically long before
graphene's fabrication in 2003 \cite{early_universal}. In the visible region
of the electromagnetic spectrum, the absorption coefficient and
transmittance of graphene have been measured experimentally and the
universal conductance has been confirmed\cite{gus,kuz,nair}.

The optical conductivity of graphene outside the low energy Dirac regime has
been calculated theoretically\cite{Chao-Ma, peres_OC}. Outside the Dirac
regime the band bending results in an increased density of states. As a
result the optical response increases and reaches a sharp maximum at the van
Hove point of $\epsilon = 2t$ where $\epsilon$ is the electronic energy and $%
t \approx 3$eV is the hopping bandwidth. The nearly total transparency of
graphene can be partially alleviated in the case of bilayer graphene\cite%
{OCBLG}. The problem is even further alleviated in the case of single layer
graphene nanoribbons in a magnetic field where the conductance can be as
much as two orders of magnitude higher than that for graphene\cite{liu1},
and recently it was shown that a subclass of bilayer nanoribbons is
similarly active in the THz-FIR regime even without a magnetic field \cite%
{OCBLGNR}.

Graphene was predicted and later experimentally confirmed to undergo
metal-semiconductor transition when fully hydrogenated (graphane).
Hydrogenated graphene loses the symmetry of A and B sublattices. It can
posses magnetic order and becomes ferromagnetic\cite{zhou}. Graphene systems
with broken inversion symmetry are of direct experimental interest.
Observation of a band gap opening in epitaxial graphene has been reported%
\cite{zhou1}. This is a direct consequense of the inversion symmetry
breaking by the substrate potential\cite{gwo}. A scheme to detect valley
polarization in graphene systems with broken inversion symmetry was
demonstrated\cite{niu}. In this paper, we shall show that due to the
inequivalence of the two valleys, the universal conductance breaks down at
low frequencies. If the on-site energy difference of the two sublattice is $%
\Delta$, absorption is only possible for frequencies above $\Delta$. The
conductance at the absorption edge is twice the universal conductance, drops
slowly as $1/\omega^2$. Therefore the universal conductance is only
approximately valid in a narrow regime overlapping with the visible band.

In the tight-binding approximation, the Hamiltonian for the graphene can be
written as,
\begin{equation}
H=\left(
\begin{array}{cc}
-t^{\prime }\alpha \left( \mathbf{k}\right) -\frac{\Delta }{2} & -tH_{12} \\
-t^{\ast }H_{12}^{\ast } & -t^{\prime }\alpha \left( \mathbf{k}\right) +%
\frac{\Delta }{2}%
\end{array}%
\right)
\end{equation}%
where $H_{12}=1+e^{i\mathbf{k}\cdot \mathbf{a}_{1}}+e^{i\mathbf{k}\cdot
\mathbf{a}_{2}}$, $\left\vert t\right\vert =2\hbar v_{F}/\sqrt{3}a,t^{\prime
}=\gamma \left\vert t\right\vert $ is the next nearest neighbor coupling,
and $\alpha \left( \mathbf{k}\right) =2\left[ \cos \mathbf{k}\cdot \mathbf{a}%
_{1}+\cos \mathbf{k}\cdot \mathbf{a}_{2}+\cos \mathbf{k}\cdot \left( \mathbf{%
a}_{1}-\mathbf{a}_{2}\right) \right] $. $\mathbf{a}_{1}$ and $\mathbf{a}_{2}$
are unit vectors given as $\mathbf{a}_{1}=a\left( \sqrt{3}/2,1/2\right) ,%
\mathbf{a}_{2}=a\left( \sqrt{3}/2,-1/2\right) $.

In the Hamiltonian, the on-site energies of the A-sublattice and
B-sublattice are $-\Delta /2$ and $\Delta /2$, respectively. The eigenvalues
are%
\begin{equation}
E_{\mathbf{k},s}=-t^{\prime }\alpha \left( \mathbf{k}\right) +s\frac{1}{2}%
\sqrt{\Delta ^{2}+4\left\vert t\right\vert ^{2}\left[ 3+\alpha \left(
\mathbf{k}\right) \right] }
\end{equation}%
with $s=\pm 1$. Correspondingly, the eigenstates can be written in the form
of%
\begin{equation}
\psi _{\mathbf{k},s}\left( \mathbf{r}\right) =\xi _{\mathbf{k},s}e^{i\mathbf{%
k}\cdot \mathbf{r}}
\end{equation}%
with%
\begin{equation}
\xi _{\mathbf{k},s}=\frac{1}{\sqrt{2}}\sqrt{1+s\frac{\Delta }{\sqrt{J\left(
\mathbf{k}\right) }}}\left(
\begin{array}{c}
-\frac{2tH_{12}}{\Delta +s\sqrt{J\left( \mathbf{k}\right) }} \\
1%
\end{array}%
\right)
\end{equation}%
where $J\left( \mathbf{k}\right) =\Delta ^{2}+4\left\vert t\right\vert ^{2}%
\left[ 3+\alpha \left( \mathbf{k}\right) \right] $. The velocity operator
can be obtained by $\widehat{\mathbf{v}}=\left( 1/\hbar \right) \left(
\partial H/\partial \mathbf{k}\right) $ and $\Delta $ is independent of $%
\mathbf{k}$,
\begin{equation*}
\widehat{\mathbf{v}}=\frac{1}{\hbar }\left(
\begin{array}{cc}
t^{\prime }\mathbf{B}\left( \mathbf{k}\right) & -itC\left( \mathbf{k}\right)
\\
it^{\ast }C^{\ast }\left( \mathbf{k}\right) & t^{\prime }\mathbf{B}\left(
\mathbf{k}\right)%
\end{array}%
\right)
\end{equation*}%
where $\mathbf{B}\left( \mathbf{k}\right) =\nabla _{\mathbf{k}}\alpha \left(
\mathbf{k}\right) =2\left[ \mathbf{a}_{1}\sin \mathbf{k}\cdot \mathbf{a}_{1}+%
\mathbf{a}_{2}\sin \mathbf{k}\cdot \mathbf{a}_{2}+\left( \mathbf{a}_{1}-%
\mathbf{a}_{2}\right) \sin \mathbf{k}\cdot \left( \mathbf{a}_{1}-\mathbf{a}%
_{2}\right) \right] $ and $C\left( \mathbf{k}\right) =\mathbf{a}_{1}e^{i%
\mathbf{k}\cdot \mathbf{a}_{1}}+\mathbf{a}_{2}e^{i\mathbf{k}\cdot \mathbf{a}%
_{2}}$.

The second quantized current density operator is given by
\begin{equation}
\mathbf{J}=e\sum_{\mathbf{k},s,s^{\prime }}\xi _{\mathbf{k},s}^{\dag }%
\widehat{\mathbf{v}}\xi _{\mathbf{k},s^{\prime }}a_{\mathbf{k},s}^{\dag }a_{%
\mathbf{k},s^{\prime }}.
\end{equation}%
The optical conductivity can be calculated from the Kubo formula,
\begin{equation}
\sigma _{\mu ,\nu }(\omega )=\frac{1}{\omega }\int_{0}^{\infty }dte^{i\omega
t}\langle \lbrack J_{\mu }(t),J_{\nu }(0)]\rangle .
\end{equation}

\begin{figure}[tbp]
\centering\includegraphics[width=9cm]{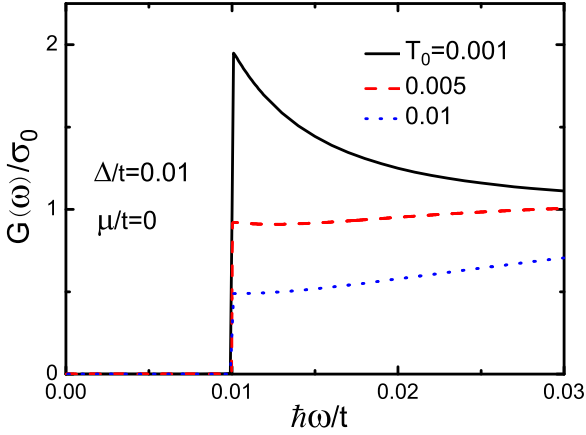} \caption{The
optical conductance in the low frequency regime, for several
different temperatures.}
\end{figure}

In Fig.1, we plot the optical conductance versus frequency for a typical
value of $\Delta =0.03eV$. The A-B on-site energy difference removes the
universal conductance which is the key feature of graphene with A-B
symmetry. For $\hbar \omega <\Delta $, the conductance is zero for any
temperature since only vertical transitions are allowed. For $\hbar \omega
>\Delta $, thermal excitation will reduce the carrier concentration near the
top of the valence band and the total interband transition rate
decreases. As a result, the conductance decreases as temperatures
increases. At the frequencies close to the energy gap, the
conductance jumps to a maximum which is greater than the universal
conductance.
\begin{equation*}
G(\omega )=\sigma _{0}\left( 1+\frac{\Delta ^{2}}{\left( \hbar \omega
\right) ^{2}}\right) \tanh \left( \frac{\beta }{4}\hbar \omega \right)
\Theta \left( \hbar \omega -\Delta \right)
\end{equation*}%
where $\sigma _{0}=\pi e^{2}/4h$ and $\beta $ is the inverse temperature in
energy units. The conductance then decreases slowly as the joint density of
states for optical transition decreases.

The low frequency conductance for difference values of $\Delta $ is shown in
Fig.2. The onset conductance decreases as $\Delta $ decreases. For systems
with complete A-B symmetry, the onset maximum disappears. In this case the
conductance starts from the universal conductance at $T=0$ and from zero at
finite temperature. The conductance at $\hbar \omega =\Delta $ is $G(\omega
)=2\sigma _{0}\tanh \left( \beta \Delta /4\right) $. Therefore, it is found
that for $T=0$, the conductance at $\hbar \omega =\Delta $ is $2\sigma _{0}$
regardless of the value of $\Delta $.

At relatively low energy around the Dirac points, $\alpha \left( \mathbf{k}%
\right) =-3+3a^{2}\left( \delta k\right) ^{2}/4$, and $E_{\mathbf{k},+}-E_{%
\mathbf{k},-}=\sqrt{J\left( \mathbf{k}\right) }$, where $\delta k$ is
measured from the $K$-point. In this region, the conductance is isotropic, ${%
Re}\sigma _{xx}^{\left( \mathbf{K}_{1}\right) }={Re}\sigma _{yy}^{\left(
\mathbf{K}_{1}\right) }=G(\omega )$, given as,
\begin{eqnarray}
G(\omega ) &=&-\frac{2\pi \hbar v_{F}^{2}\sigma _{0}}{\omega }\sum_{\delta
\mathbf{k}}\left( \tanh \frac{\beta }{2}E_{\delta \mathbf{k},+}-\tanh \frac{%
\beta }{2}E_{\delta \mathbf{k},-}\right)   \notag \\
&&\left( 1+\frac{\Delta ^{2}}{\eta }\right) \left[ \delta \left( \hbar
\omega -\sqrt{\eta }\right) -\delta \left( \hbar \omega +\sqrt{\eta }\right) %
\right]
\end{eqnarray}%
Here, $\eta \left( \delta k\right) =\Delta ^{2}+4\hbar ^{2}v_{F}^{2}\left(
\delta k\right) ^{2}$,
\begin{equation*}
E_{\delta \mathbf{k},s}=\gamma \frac{2\sqrt{3}\hbar v_{F}}{a}\left[ 1-\frac{%
\gamma }{4}a^{2}\left( \delta k\right) ^{2}\right] +s\frac{1}{2}\sqrt{\eta }
\end{equation*}

\begin{figure}[tbp]
\centering\includegraphics[width=9cm]{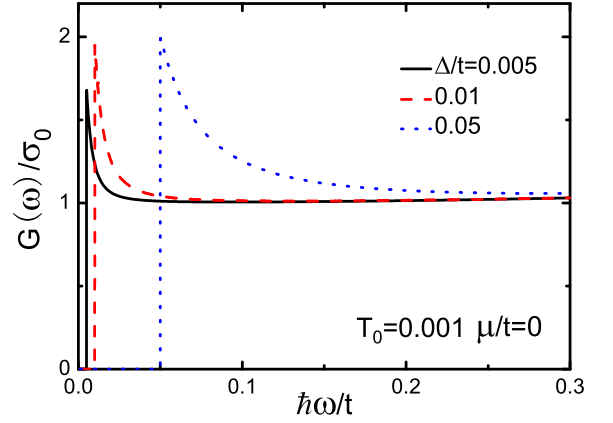} \caption{The
optical conductance in the low frequency regime, for several
different values of $\Delta$.}
\end{figure}

In Fig.3, we show the zero temperature conductance over a wide frequency
range of $0$ to $2t$. The absorption at frequencies below the gap $\Delta $
is now forbidden, as expected. The interesting feature of this systems is
the onset of absorption at the band edge. The absorption at the edge is
discontinuous and jumps vertically from zero to a value twice the universal
conductance. This behavior can be derive analytically around the Dirac
regime as follows.

In the limit of $T\longrightarrow 0$, $\tanh \frac{1}{2}\beta E_{\mathbf{k}%
,+}-\tanh \frac{1}{2}\beta E_{\mathbf{k},-}\simeq \Theta \left( \Delta
^{2}+4\hbar ^{2}v_{F}^{2}\left( \delta k\right) ^{2}\right) $, we obtain a
simple form for the conductance due to the contribution from the 6 Dirac
points, given as $G(\omega )=\sigma _{0}\left[ 1+\Delta ^{2}/\left( \hbar
\omega \right) ^{2}\right] \Theta \left( \hbar \omega -\Delta \right) $. At $%
\hbar \omega =\Delta +0^{+}$ for any $\Delta $, the conductance is twice of
the universal conductance. In other words, for graphene with A-B asymmetry,
while the universal conductance is broken, the onset conductance at the band
edge for $T=0$ has a universal value of $2\sigma _{0}$. In comparison to the
wellknown interband absorption rate in semiconductors, $G_{semicond}(\omega
)\sim \sqrt{\hbar \omega -\Delta }$, this feature of interband absorption in
graphene is very unique. At higher frequency, the conductance increases with
frequency. Therefore an inequivalence of A-B sublattices can remove the
universal conductance, or limit its applicability to a small regime around
the visible frequency.

\begin{figure}[tbp]
\centering\includegraphics[width=9cm]{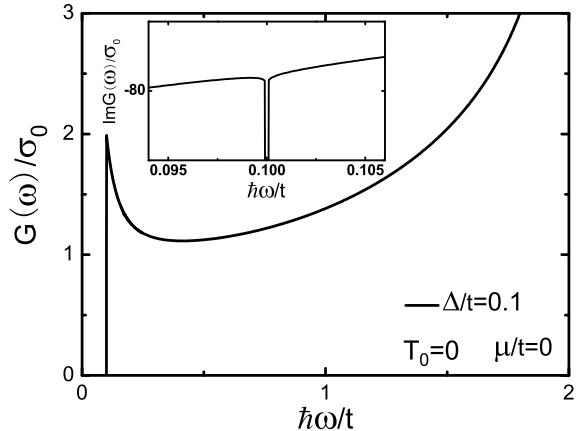} \caption{The
frequency dependent optical conductance, $G(\protect\omega)$. It
is expressed in unit of universal conductance.}
\end{figure}

The inset of Fig.3 shows the imaginary part of the conductivity. Near the
band edge, the imaginary part is very large which indicate phase difference
of nearly $\pi/2$ between the incident field and the current response. This
phase difference decreases rapidly as frequency increases. The size of the
dip at $\hbar\omega=\Delta$ is about the same order of magnitude as the jump
in the conductance.

The total integrated absorption is given by
\begin{equation}
\sigma_{total} = \int d\omega G(\omega).
\end{equation}
Experimentally\cite{nair} it has been shown that for graphene with A-B
symmetry, the universal conductance is valid up to the high end of the
visible band, $\lambda=400nm$ or $\hbar\omega=2eV$. Theoretically it is
shown that the conductance is approximately $\sigma_0$ up to $\hbar\omega=t$%
. Therefore for graphene with A-B symmetry, we integrate (10) from 0 to $t$
and $\sigma_{total}=2t\sigma_0$. For graphene without A-B symmetry, we
integrate (10) from $\Delta$ to $t$ and $\sigma_{total}=2t\sigma_0(1-%
\Delta^2/t^2)$. Since $\Delta^2/t^2 \approx 0.01$, we can conclude that the
total optical response is nearly conserved even when a gap opens up at the
Fermi energy. For small $\Delta$, any loss of optical response due to the
gap is recovered in the regime of $\hbar\omega$ slightly higher than $\Delta$%
. For large $\Delta$, the region required for recovering the optical
response is wider. This can be seen quantitatively in Fig.2.

\begin{figure}[tbp]
\centering\includegraphics[width=9cm]{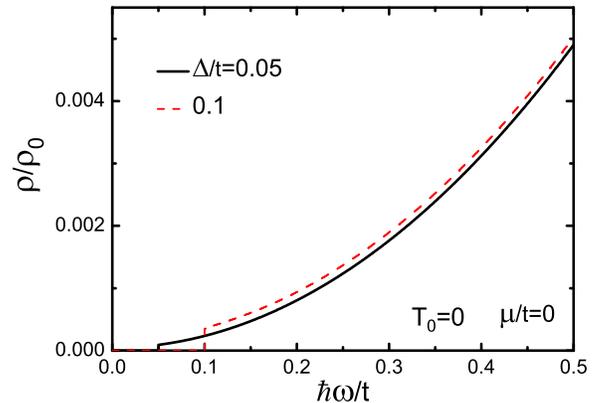} \caption{The
frequency dependent resistivity, $\protect\rho(\protect\omega)$.
It is expressed in unit of in verse universal conductance
$\protect\rho_0$. }
\end{figure}

Finally we show in Fig.4 that the resistivity of the system,
defined as $ \rho(\omega)=Re[1/\sigma(\omega)]$, has a well
behaved monotonic frequency dependence. Apart from a small jump at
the band gap, the resistivity varies with frequency approximately
parabolically below the van Hove point.

In conclusion, we have shown that for graphene without inversion symmetry
the optical conductance becomes strongly frequency dependent. However, the
onset conductance at the band edge has a universal value, regardless of the
value the band gap.

\textit{Acknowledgement}-- This work is supported in part by the National
Natural Science Foundation of China and the Australian Research Council.

\bigskip \noindent{$^{a)}$Electronic mail: leichen@pku.edu.cn}\newline
\bigskip \noindent{$^{b)}$Electronic mail: czhang@uow.edu.au}\newline

\end{document}